\documentstyle[aps,preprint]{revtex}
\renewcommand{\baselinestretch}{1.35}
\begin{document}
\title
{Eigenvalue problems for the complex PT-symmetric Potential $V(x)=igx$}
\author {Zafar Ahmed\\
Nuclear Physics Division, Bhabha Atomic Research Centre,\\
Mumbai 400 085\\
zahmed@.barc.gov.in}
\maketitle
\date{}
\begin{abstract}
The spectrum of complex PT-symmetric potential, $V(x)=igx$, is known to be
null. We enclose this potential in a hard-box: $V(|x| \ge 1) =\infty $ and
in a soft-box: $V(|x|\ge 1)=0$. In the former case, we find real discrete
spectrum and the exceptional points of the potential. The asymptotic
eigenvalues behave as $E_n \sim n^2.$ The solvable purely imaginary
PT-symmetric potentials vanishing asymptotically known so far do not have
real discrete spectrum. Our solvable soft-box potential possesses two real
negative discrete eigenvalues if $|g|<( 1.22330447)$. The soft-box potential
turns out to be a scattering potential not possessing reflectionless states.
\\
PACS NO. : 03.65.Ge
\end{abstract}
\vskip 1.cm
A complex Hamiltonian that does not change under the joint transformation of
Parity (P: $x \rightarrow -x$) and Time-reversal ($i \rightarrow -i$) is
called PT-symmetric: (PT)H(PT)$^{-1}=$H or [PT,H]=0. When two (linear)
operators A and B commute, we can arrage for  simultaneous eigenstates. However, if
one of the operators is anti-linear: A $(c \psi)=c^* \psi^*$, then for A and B
we will have simultaneous eigenstates if only the eigenvalues of B are real.
In case of complex eigenvalues of B, the simultaneity of eigenstates does
not hold. Further, it follows that B can have real or complex conjugate pairs
of eigenvalues. Therefore, PT being anti linear a Hamiltonian commuting with
it may have entire discrete spectrum as real or as complex conjugate pairs
or both. In the first case PT-symmetry is unbroken or exact in the second
case it breaks down spontaneously.
\par The revelation that a complex PT-symmetric potential [1] too may possess
real discrete spectrum has lead to very interesting developments [2].
In this regard, the exactly solvable potentials have been very revealing [3].
Quasi-exactly solvable complex PT-symmetric potentials [4] have also been
helpful wherein one gets only a certain number of eigenvalues and eigenstates
witb some restriction on the parameters of the potential.
Next some simple semi-analytically [5] solvable potentials have been proposed wherein
one gets implicit equation for energy to calculate the discrete spectrum
iteratively and numerically.
These potentials have been useful and handy in some tedious calculations
\par For the potential $V_B(x)=-(ix)^\nu$ [1], a few critical values of $\nu$
in (1.42207,2) are known where the spectrum has a few real eigenvalues and
for $\nu \ge 2$  the entire spectrum is real. For analytically solvable
potential, $V_S(x)=-V_1 \mbox{sech}^2 x+iV_2 \mbox {sech}x \tanh x$, [3] the
discrete spectrum is real if $|V_2|$ is less than or equal to a critical
value $\sqrt{V_1+1/4}$. Otherwise the spectrum consists of complex conjugate
pairs of energy. The semi-analytically solvable potential $V_Z(|x|<1)=iZ
\theta(x),V_Z(|x|>1)=\infty$ [5] has been found to have critical values of
Z above which PT-symmetry is spontaneously broken. Here $\theta(x \le 0)=-1$
and $\theta(x>0)=1$.
\par Nowadays, with the advent of the packages like {\it Mathematica} one can
perform quick and almost error-free calculations involving even the higher
order functions. This gives us a broader scope to extend at least the class of
semi-analytically solvable potentials. In this Letter, we present
the complex PT-symmetric potential $V(x)=igx$ in two modifications.
These two models are semi-analytically solvable which also arise due to
the following open questions in PT-symmetric quantum mechanics [6,7].
\par Another way of visualizing real spectrum of a non-Hermitian Hamiltonian
is through the exceptional points of a Hamiltonian as proposed by Kato (1960)
[8,9]. Let the eigenvalues $E_n(\lambda)$, of $H_\lambda=H_1+\lambda H_2$ be
analytic function of $\lambda$, where $H_1, H_2$ are Hermitian Hamiltonians.
If $\lambda$ is analytically continued say as $\lambda=i\mu$ in a complex
domain, then at certain values of $\mu=\mu_c$ two real distinct eigenvalues
and eigenstates may coalesce (merge) in to one. Further, if $\mu>\mu_c$
the coalesced eigenvalues reappear as complex conjugate pair. In this
complex extension of the parameter $\lambda$ the Hamiltonian becomes
non-Hermitian. We strongly feel that PT-symmetry or more generally
pseudo-Hermiticity [10] ($\eta H \eta^{-1}=H^\dagger$) may be the necessary
condition for the existence of exceptional points  in H$_{i\mu}$.
\par So far we do not know the necessary and sufficient condition to
determine {\it a priori} whether a given complex PT-symmetric potential will
have real discrete spectrum. However, case of $V(x)=igx$ is well studied
wherein using the arguments of Stokes and Anti-stokes lines the
spectrum has been found to be null. An interesting {\it a priori} and {\it necessary}
condition in this regard is that the potential ought to have a pair of classical
turning points (solutions of $E=V(x)$) of the type $\pm a+ib$ [11] at a given
real energy. In addition to this, any singularity of the potential should not
lie on the line from $z=-a+ib$ to $z=a+ib$.
This explains why the real discrete spectrum of $V_L(x)=igx [1], V_E(x)=iV_2
\tanh x \mbox {sech} x$, $V_G(x)=iV_2xe^{-x^2}$ and $V_C(x)=iV_2x/(1+x^2)$
for {\it any real value of} $V_2$ is empty. Since only $V_E(x)$ among these potentials is analytically solvable
having the property of shape-invariance, therefore an argument based on
supersymmetry may be ruled out for the emptiness of the real discrete spectrum.
However, when a real attractive potential of sufficient strength is added,
these potentials [3] do possess real discrete spectrum (e.g., $V_{HO}(x)=x^2+igx$).
\par In this work, we study the purely imaginary potential, $V_L(x)$, within
a hard-box ($(V(|x|>1) = \infty)$ or a soft-box $(V(|x|\ge 1)=0)$ for real
discrete spectrum.
The abovementioned potentials, $V_G(x)$ and $V_E(x)$ may also be seen as purely
imaginary PT-symmetric potentials vanishing asymptotically and possessing
the real discrete spectrum as null. Similarly, we find that simple purely
imaginary potential $V(x)=iV_0\theta(x)$ placed within the soft box
also has real discrete spectrum as null. This gives us an additional
motivation to discuss $V_L(x)$ placed within a soft box.
\par First we study
\begin{equation}
V(x)=igx, V(|x|\ge 1)=\infty.
\end{equation}
Taking $2m=1=\hbar^2$ in one-dimensional time-independent  Schr{\"o}dinger
equation, we get the wave solution as $\psi(x)={\cal A} Ai(z)+ {\cal B} Bi(z)$
with $z=(E-igx)/q, q=(g^2)^{1/3}$. The Dirichlet boundary conditions that
$\psi(x=\pm 1)=0$ and their simultaneous consistency gives rise to an implicit
and transcendental equation for $E$  as
\begin{equation}
Ai[(E+ig)/q] Bi[(E-ig)/q]-Ai[(E-ig)/q] Bi[(E+ig)/q]=0
\end{equation}
Here $Ai(z)$ and $Bi(z)$ are the Airy functions. The eigenfunction is expressible
as
\begin{equation}
\psi(x)={\cal C}(E,g) Ai[(E+igx)/q]+Bi[(E+igx)/q],~~~-1 \le x \le  1,
\end{equation}
where
\begin{equation}
-{Bi[(E+ig)/q] \over Ai[(E+ig)/q]} = {\cal C} (E,g) = -{Bi[(E-ig)/q]
\over Ai[(E-ig)/q]}.
\end{equation}
For the real branch of $q$, the Eq. (2) say $f(E)=0$ is a real equation
of $E$ on real line. Therefore as per the {\it fundamental theorem of algebra}
the roots (eigenvalues) will be either purely real or complex-conjugate
pairs. In the former case, $\psi(x)$ will also be an eigenstate of the
anti-linear operator PT ($x\rightarrow -x$ and $i\rightarrow -i$). In the
latter case, $\psi(x)$ will no more be the eigenstates of PT and the
PT-symmetry will be spontaneously broken.
We employ {\it Mathematica} to perform the calculations. We find the first
critical value $g_1=12.3124556046$ of $g$. If $|g|<g_1$, each calculated
eigenvalue (root of from Eq. (2)) is real and discrete. Then if $|g|=g_1$,
we find that only the lowest pair of real discrete eigenvalues merge
(coalesce) into one real eigenvalue to make way (when $|g|>g_1$) for a single
pair of complex conjugate eigenvalues signifying spontaneous breakdown of
PT-symmetry. When $|g|=g_1$, we notice the merger of lowest real eigenvalues
at $E=7.1086$. A sudden jump discontinuity is observed in the ground level
while crossing a critical value e.g., $E_G(12.31)=7.0316$, $E_G(12.32)=
21.7209$. The first five eigenvalues for $g=12.31$ are 7.03165, 7.1848,
21.7217, 39.1884, 61.4929. On the other hand, the first five eigenvalues
for $g=12.32$ are $7.1097\pm .1342i$, 21.7209, 39.1880, 61.4926(61.6850).
The number in bracket equals $25 \pi^2/4$. The next critical values of $g$
are $g_2\approx 53.18,g_3\approx 122.90 ,...$ The first five eigenvalues
for $g=53$ are $16.4942 \pm 24.4311i$, 29.7350, 31.6153, 58.2078 and for
$g=54$ they are $16.7009 \pm 25.0725 i, 30.8513 \pm 1.9731 i$ and 58.0801.
\par Thus each critical value of $g$ indicates the onset of the removal of
lowest pair of {\it real} discrete eigenvalues. Removal of only ground level of a
potential from its supersymmetric partner potential has earlier been observed
in supersymmetric quantum mechanics [12]. In contrast to this, a complex
PT-symmetric potential having infinite real discrete spectrum manifests
in the removal of the lowest pair of {\it real} eigenvalues as its parameter passes through several critical values.
When this happens the ground state eigenvalue becomes discontinuous and we
have $\Delta(g_1)=[(E_G(|g|=g_1+\epsilon)-E_G(|g|=g_1-\epsilon)]$. When the
parameter $g$ increases the net density of states of {\it real} eigenvalues reduces.
\par Seeing the potential (1) as a non-Hermitian PT-symmetric perturbation
to the one dimensional box having $E^0_n=n^2\pi^2/4$ presents interesting
features. We find that there are two remarkable features of the complex
PT-symmetric perturbation (1) in contrast to its real Hermitian counterpart
when $V(|x|<1)=gx$. In the former case, as the value of $g$ increases the
lowest eigenvalues(approximately corresponding to $E^0_n$) of the real
discrete spectrum get removed reducing the number of {\it real} eigenvalues
below a fixed energy $E=E^\ast$. On the other hand, in the Hermitian case, these lowest
eigenvalues get perturbed and get pushed down to become negative however
the number of levels below $E=E^\ast$ does not change. Secondly, we find
that the higher real eigenvalues behave as $E_n \sim E^0_n+\epsilon^2_1$
whereas in the Hermitian case they are like $E_n=E^0_n-\epsilon^2_2$,
$\epsilon_1, \epsilon_2$ are small real numbers.
\par We also find that there are bands of values of $g$, wherein for a fixed value of
$g$, we get a continuous band of energies which satisfy the eigenvalue equation
(2), however the corresponding eigenstates are null. The discrete eigenvalues
do exist on the left and right of this energy-band. For example when $g=3.4$,
we get a continuous band of energies $23.2 < E < 26.5$  wherein every energy
is an ``eigenvalue'' with a null eigenvector. For $g=5.0 i$, we get such
an energy band as $ 32.8 < E < 34.2$.
\par The discrete spectrum of the Hamiltonian with a hard-box potential
can also be conveniently calculated by diagonalizing the matrix $H_{r,s}=
<r|H|s>$, where $|r>$ are the complete orthnormal eigenstates of the
one-dimensional hard-box potential namely $V_H(|x|<1)=0, V_H(|x|>1)=\infty$.
We get for our hard-box potential
\begin{equation}
H_{r,s}={r^2\pi^2 \over 4}\delta_{r,s}+(-1)^{t/2}{16i g r s \over
\pi^2 (r^2-s^2)^2}[{1+(-1)^{t} \over 2}],~~ t=r+s+1, r,s=1,N.
\end{equation}
Here $N$ denotes the truncation or the order of the ``infinite'' dimensional
matrix $H_{r,s}$. The eigenvalue determinant $\det |H-EI|_{N \times N}$
is a very fast converging function of $N$. Consequently, the critical
values of $g$ discussed above hardly depend upon $N$. This in turn also
means that first $N$ eigenvalues (real or complex) calculated by fixing
a value of $g$ and the dimensions of the $H$ matrix $N \times N$ would
be quite accurate. We would like to remark that the method of diagonalization
of eigenvalues supplements the calculations using Eq. (2) specially when
eigenvalues are complex conjugate pair.
\par The PT-symmetry is also conceived as P-pseudo-Hermiticity [9], it is
therefore worth pointing that the matrix $H_{r,s}$ (5) for the PT-symmetric
matrix is pseudo-Hermitian for any value of $g$ with parity operator given
by $P=(-1)^r \delta_{r,s}.$ Other property of $H_{r,s}$ is that it is
complex symmetric matrix. The experience gained in the diagonalization of
a hard-box Hamiltonian is markedly different from that of diagonalization
of the well known Hamiltonian of D. Bessis $H_{DB}=p^2-igx^3$ [1].
\par The diagonalization of $H_{DB}$ in the real harmonic oscillator basis
by truncating the dimensions of the matrix to finite:  $N \times N$,
contrary to yielding the whole discrete spectrum as real for any value of $g$
[1], yields  [13] a finite number ($N$) of eigenvalues as real if
$|g| \ge g_c(N)$. Here $g_c(N)$ is crucially decreasing function of $N$.
So eventually as $N \rightarrow \infty$, $g_c(N)$ vanishes justifying
entire spectrum to be real for any real value of $g$.
\par The complex PT-symmetric Potential
\begin{equation}
V_R(|x|<1)=-V_1+iV_2, V_R(|x|>1)=0
\end{equation}
possesses scattering states for $E>0$ and bound states for $E<0$.
The interesting scattering properties of (6) have earlier been studied
for this potential [14]. We further find that when $V_1=0$, the purely
imaginary potential (6) has the real discrete spectrum as empty.
This potential along with $V_E(x)$ and $V_G(x)$ [3] constitute purely
imaginary and asymptotically vanishing PT-symmetric potentials that do
not possess any real discrete eigenvalue. We now study the second
modification of $igx$-potential
\begin{equation}
V(|x|<1)=igx, V(|x|>1)=0,
\end{equation}
the soft-box potential that is exceptional in this regard.
Assuming $2m=1=\hbar^2$, we consider the Schr{\"o}dinger equation for (7)
when $E<0$. Using the Airy functions again, we write the wave solution of
Schr{\"o}dinger equation in three regions as $\psi(x<-1)=c e^{\kappa x},
\psi( -1 \le x \le 1)=a Ai(z)+ b Bi(z), \psi(x>1)=d e^{-\kappa x}.$ Matching
the wave solution at $x=\pm 1$ and eliminating $a,b,c,d$, we get
\begin{equation}
[\kappa q Ai(s) + ig Ai^\prime (s)][\kappa q Bi(t) - ig Bi^\prime (t)] =
[\kappa q Bi(s) + ig Bi^\prime (s)][\kappa q Ai(t) - ig Ai^\prime (t)]
\end{equation}
Here $s,t=(E \pm ig)/q, z=(E-igx)/q, q=(g^2)^{1/3}$ and $\kappa=\sqrt{-E}$.
This equation is an implicit equation of $-E$ which is real for real
negative values of E. So as per the {\it fundamental theorem of Algebra}
it will have either negative real roots or complex conjugate roots.
By the evaluation of the roots of Eq. (8), we find that if $|g|
< 1.223830447=g_c$ there exist two negative real discrete eigenvalues.
Further if $|g|=g_c$, we find that two real discrete eigenvalues
merge at $E=.24994$. For the values of $|g| > g_2,$ the eigenvalues
are complex conjugate pairs. As $|g|$ is decreased one level is pushed down
to negative values, the other one shifts closer to zero. For instance for
$|g|=1.2$, $E_1=-0.40891, E_2=-.14426$. For $|g|=0.6$, we get $E_1=-2.58012,
E_2= -0.00275$. When $|g|=0.1$, we get $E_1= -9.29466, E_2=-1.7849 \times
10^{-6}$.
We conjecture that the PT-symmetric potentials which have both bound
and scattering states will have only one critical value of parameter as
found here for (7) and in the models in Refs. [3].
\par The limit when $g$ approaches zero is not transparent
here. For this we use the asymptotics or the Airy functions $Ai(s),Bi(s)
\sim s^{1 \over 4} e^{\mp 2s^{3/2}/3}$ for large values of $s \approx t
(\sim {E \over g^{2/3}})$. We find that the Eq. (8) becomes independent of
energy and it is trivially satisfied for $g=0$. This confirms the Hermitian
free particle limit of the soft-box potential wherein any  energy
is a solution of Eq. (8) and hence discrete energies do not exist.
\par Thus, we have at least one purely imaginary PT-symmetric potential
vanishing asymptotically and possessing real discrete spectrum. This
investigation receives significance due to interesting results [15] of
Hermitian quantum mechanics like: a real potential well such that
$\int_{-\infty}^{\infty} V(x) dx < \infty$ has at least one discrete eigen
value irrespective of the depth and width of the potential. In the absence
of potentials like (7), it would be as though a purely imaginary PT-symmetric
potential that vanishes asymptotically does not possess real discrete spectrum.
\par Earlier we have proposed generation of discrete eigenvalues of a Hermitian
 potential $V(x)=-x^4$ [16] by imposing PT-symmetric boundary condition on the wave
 function. At these discrete energies the reflection probability vanishes and
 potential becomes reflectionless [16,17].
If we analytically continue $\kappa$ as $\kappa=ik$ in the eigenvalue
equation (8) we get
\begin{equation}
[k q Ai(s) + g Ai^\prime (s)][k q Bi(t) -g Bi^\prime (t)] =
[k q Bi(s) + g Bi^\prime (s)][k q Ai(t) -g Ai^\prime (t)], k =\sqrt{E}, E>0.
\end{equation}
This equation is complex on the real line of $E$, so according to {\it fundamental
theorem of algebra} it will  have complex roots and some of these roots may
or may not be real. For $E>0$, the states are scattering states therein real
discrete eigen value/ values if occur would be embedded in positive energy continuum.
These states are discrete energy eigenstates generated by imposing
PT-symmetric boundary conditions as $\psi(\pm x)\sim e^{\pm ikx},k=\sqrt{E},
E>0$ instead of the usual $\psi(\pm x) \sim e^{\pm \kappa x}, \kappa=
\sqrt{-E}, E<0.$ However, for our soft-box potential (7) we  have carried
out a very careful search and we do not find real solution(s) of the Eq. (9).
Therefore, such reflectionless states do not seem to occur for this potential.
\par One interesting common feature of $gx$-potentials lies in the presence of
$q=(-g^2)^{1/3}$(cube root of $-g^2$) which has one real ($q_0$) and two complex
$(q_1,q_2)$ branches. For the Hermitian case, we find that all the three branches
of $q$ give identical discrete spectrum. However for a given eigen value, we
have three eigenstates $\psi_{q_0}(x),\psi_{q_1}(x)$ and $\psi_{q_2}(x)$.
The first one is real and other two are complex valued functions of $x$.
Since there can be no degeneracy in one-dimension, we therefore find that the
last two eigenstates differ from the first one by multiplicative complex
constants only. For the non-Hermitian case of $igx$-potentials
(1,7), we find that the complex branches $q_{1,2}$ may not yield the full
real discrete spectrum  as generated by the main (real) branch $q_0$.
Only first few eigenvalues of the main branch are repeated. Amusingly,
the  continuous band of the higher energies turns out to be ``eigenvalues''
however with null eigenvectors. Every repeated real eigenvalue for the three
branches once again yield three corresponding complex eigenfunctions:
$\psi_{q_0}(x),\psi_{q_1}(x)$ and $\psi_{q_2}(x)$. Only the first one is also
eigenstate of PT, whereas other two are not. We once again find the last two eigenstates differ from
the first one by a complex multiplicative factor. This wards off two fake
occurrences:  degeneracy (in one dimension) and the breaking PT-symmetry
(despite real eigenvalues) when one chooses a complex branch of $q$.
The complex branches of $q$ do not create any new energy eigenvalue
other than those yielded by the real branch.
\par Finally, we conclude that though the spectrum of $V(x)=igx$ is null,
however, the spectrum of two of its modifications are interesting.
These are semi-analytically solvable models. The PT-symmetric hard-box
potential helps studying issues like exceptional points, reality of eigen
values, spontaneous PT-symmetry breaking and some interesting
features of  Hamiltonian-diagonalization. The reduction in the density of
{\it real} eigenvalues as the parameter $g$ increases is found to be disparate
and strong feature of PT-symmetry. We conjecture that PT-symmetry or
pseudo-Hermiticity could be a necessary condition for observing exceptional
points in the analytically continued complex Hamiltonian. Our soft-box
potential is the first solvable model purely imaginary PT-symmetric which vanishes
asymptotically and has real discrete spectrum. In the light of this new possibility,
it could further be interesting to search for purely imaginary and asymptotically
vanishing potentials having discrete spectrum. It may be that the higher-order
imaginary potentials
of the type $V_n(x)=iV_2 \mbox{(sech)}^n x \tanh x, n>1$ do possess real discrete
spectrum since they will have PT-symmetric turning points of the type $\pm a+ib$
Since these are not analytically solvable it would require numerical calculations.
\par We do not find reflectionless states embedded in positive
energy continuum in our soft-box potential.
\par Very importantly, it turns out that all the three mathematical
branches of $q$ altogether output an unambiguous and non-anomalous physical
result (spectrum) in both the cases whether we have Hermitian or PT-symmetric
Hamiltonians. Perhaps only $gx$ or $igx$ potentials that are amenable to
analytic solutions facilitate such a study. This study has not been without
interesting surprises as mentioned earlier. Just recently both $gx$ and $igx$
potentials have been found very interesting for their complex classical
trajectories [18].
\par
We wish that the two modifications of $igx$-potential presented here will
also be helpful in understanding other issues of PT-symmetric quantum
mechanics in future.

\section*{References}
\begin{enumerate}
\item C.M. Bender and S. Boettcher, Phys. Rev. Lett. {\bf 80} (1998) 5243.
\item e.g., see articles and Refs. in The special issue, J. Phys. A: Math. Gen.
{\bf 39} (2006) 9965-10261.
\item Z. Ahmed, Phys. Lett. A {\bf 282} (2001) 343.
\item
C. M. Bender and S. Boettcher, J. Phys. A: Math. Gen. {\bf 31} (1998) L273. \\
A. Khare and B.P. Mandal, Phys. Lett. A {\bf 272} (2000) 53. \\
      B. Bagchi, S. Mallik, C. Quesne and R. Roychoudhury, Phys. Lett. A
      {\bf 289} (2001) 34.
C. M. Bender, M. Monou, J. Phys. A: Math. Gen. {\bf 38} (2005) 2179.
\item M. Znojil and G. Levai, Mod. Phys. Lett. A {\bf 16} (2001) 2273.
\item G.S. Japaridze, J. Phys. A: Math. Gen. , {\bf 35} (2002) 1709.
\item Z. Ahmed, Phys. Lett. A {\bf 360} (2006) 238.
\item T. Kato, {\it Perturbation theory of linear operators}, Springer,
          Berlin (1966).
\item W.D. Heiss, J. Phys. {\bf 54} (2004) 1091.
\item E.C.G. Sudarshan, Phys. Rev. {\bf 123} (1961) 2183.\\
      T.D. Lee and G.C. Wick, Nucl. Phys. B{\bf 9} (1968) 209.\\
      A. Mostafazadeh, J. Math. Phys. {\bf 43} (2002) 205.
\item Z. Ahmed, J. Phys. A {\bf 38} (2005) L301.
\item F. Cooper, A. Khare and U. P. Sukhatme,Phys. Rep. {\bf 251} (1995) 267.
\item C.K. Mondal, K. Maji, S.P. Bhattacharyya, Phys. Lett. A,
{\bf 291} (2002) 307
\item Z. Ahmed, Phys. Lett. A {\bf 324} (2004) 152.
\item M. Reed and B. Simon, {\it Methods in Modern Mathematical Physics:
III Scattering Theory} (Academic Press, New York) 1970.
\item Z. Ahmed, C.M. Bender and M. V. Berry, J. Phys. A: Math. Gen. {\bf 38 }
 (2005) L 627.
\item Z. Ahmed, J. Phys. A: Math. Gen. {\bf 39} (2006) 7341.
\item C. M. Bender, D. D. Home and D.W. Hook,  quant-ph/0609068
\end{enumerate}
\end{document}